# Lung Segmentation in Chest X-rays with Res-CR-Net

**Haikal Abdulah[1,4‡], Benjamin Huber[1,4‡], Sinan Lal[1], Hassan Abdallah[3], Hamid Soltanian-Zadeh[4], Domenico L. Gatti[1,2*]**

[1]Department of Biochemistry, Microbiology and Immunology, Wayne State Univ., Detroit, MI, USA
[2]NanoBioScience Institute, Wayne State Univ., Detroit, MI, USA
[3]Department of Biostatistics, University of Michigan, Ann Arbor, MI, USA
[4]Departments of Radiology and Research Administration, Henry Ford Health System, Detroit, MI, USA

‡These authors, listed in alphabetical order, contributed equally to the study.

*E-mail: dgatti@med.wayne.edu

## Abstract

Deep Neural Networks (DNN) are widely used to carry out segmentation tasks in biomedical images. Most DNNs developed for this purpose are based on some variation of the encoder-decoder U-Net architecture. Here we show that Res-CR-Net, a new type of fully convolutional neural network, which was originally developed for the semantic segmentation of microscopy images, and which does not adopt a U-Net architecture, is very effective at segmenting the lung fields in chest X-rays from either healthy patients or patients with a variety of lung pathologies.

## 1. Introduction

Chest X-rays (CXRs) are still among the most used imaging tests worldwide [1], with millions of CXRs generated annually. The correct interpretation of CXRs is a major challenge to radiologists, and researchers continue to develop methods of machine learning (ML) and Artificial Intelligence (AI) to assist radiologists in this task [2]. An important component of these methods is the process of semantic segmentation [3], which provides a mask of the lung regions, and by exclusion of the non-lung region. Automated segmentation of the lungs in CXRs is a challenging problem due to the presence of strong edges at the rib cage and clavicle, the lack of a consistent lung shape among different individuals, and the presence of cardiovascular structures in the chest. When patients are healthy, because of the high contrast between the lung fields and their boundaries, ML/AI methods usually provide reliable segmentations. However, when patients have pathologies that produce lung density abnormalities, the lower contrast between the lungs and the surroundings makes the segmentation task in CXRs significantly more challenging [4].

Deep learning/Neural Network (DL/NN) approaches have become very popular to address these challenges, with the widespread use of encoder-decoder based convolutional neural networks (CNN) to perform lung segmentation (reviewed in [5, 6]). These method were typically tested with the Japanese Society of Radiological Technology (JSRT, [7, 8]), the Montgomery County (MC), and the Shenzen Hospital (SH) ([4, 9], and https://www.kaggle.com/yoctoman/shcxr-lung-mask) datasets of segmented lungs. More recently, Carvalho Souza *et al.* [10] have proposed a method to incorporate pathology associated opacities as part of the lung segmentation by using a first NN to derive an initial segmentation that excludes the opacities, and a second NN to reconstruct the lung regions "lost" due to pulmonary abnormalities. Selvan *et al.* [11] have instead treated the high opacity regions as missing data and presented a modified CNN-based image segmentation network that utilizes a deep generative model for data imputation. Kholiavchenko *et al.* [12] have reported that further improvement in lung segmentation can be achieved by training the NN not only on the ground-truth segmentation masks but also on the corresponding contours.

Recently, we have introduced a new type of fully convolutional (FC), residual NN [13, 14], Res-CR-Net, that departs from the widely adopted of U-Net models [15-18] with encoder-decoder architecture. Res-CR-Net combines residual blocks based on separable, atrous *c*onvolutions [19, 20] with residual blocks based on recurrent NNs [21]. This network displayed excellent performance when tasked to segment images from either electron or light microscopy in three/four separate categories, using only a small number of images for training [22]. Here we show that Res-CR-Net, in the configuration used for microscopy images without additional hyperparameters refinement, achieves excellent performance also in the lung segmentation of normal and pathologic CXRs.

## 1. Methods

### 1.1 CXRs and lung segmentations sources

1. Japanese Society of Radiological Technology (JSRT) dataset [7]. The dataset consists of 247 posterior-anterior (PA) chest radiographs with and without chest lung nodules, with a resolution of 2048 × 2048, 0.175 mm pixel size, and 12-bit depth. The reference organ boundaries for the JSRT images for left and right lung fields, heart and left and right clavicles were introduced by van Ginneken *et al.* [8] in 1024 × 1024 resolution and available in the Segmentation in Chest Radiographs (SCR) database (https://www.isi.uu.nl/Research/Databases/SCR/).

2. Montgomery County (MC) dataset [9] (http://openi.nlm.nih.gov/imgs/collections/NLM-MontgomeryCXRSet.zip ). This dataset, publicly available from the Department of Health and Human Services of Montgomery County (Maryland), was collected by the Montgomery County's Tuberculosis Control program and consists of 138 CXR images, of which 80 are from normal patients and 58 are from patients with some manifestation of tuberculosis. The CXR images are available in 12-bit gray-scale with resolutions of either 4020 × 4892 or 4892 × 4020 and 0.0875 mm pixel spacing in both horizontal and vertical directions. The MC dataset has lung segmentation masks excluding the heart and large vasa, which were marked under the supervision of a radiologist and made available by Candemir *et al.* [4].

3. Shenzhen Hospital (SH) dataset [9]. X-ray images in this dataset have been collected by Shenzhen No. 3 Hospital in Shenzhen, Guangdong providence, China (http://openi.nlm.nih.gov/imgs/collections/ChinaSet_AllFiles.zip ). The X-rays were acquired as part of the routine care at Shenzhen Hospital. The set contains images in JPEG format. There are 326 normal X-rays and 336 abnormal X-rays showing various manifestations of tuberculosis. Lung segmentation masks, bounded by the heart and large vasa, were prepared manually by students and teachers of the Computer Engineering Department, Faculty of Informatics and Computer Engineering, National Technical University of Ukraine "Igor Sikorsky Kyiv Polytechnic Institute", Kyiv, Ukraine [23].

4. V7-Darwin dataset (https://darwin.v7labs.com/v7-labs/covid-19-chest-x-ray-dataset). This dataset contains 6500 images of AP/PA chest X-rays with pixel-level polygonal lung segmentations. 5863 images are sourced from https://data.mendeley.com/datasets/rscbjbr9sj/2 (also available and commonly referred to by the Kaggle dataset: https://www.kaggle.com/paultimothymooney/chest-xray-pneumonia/data). In addition to several images of normal lungs, the dataset includes 1970 images of viral pneumonia, 2816 images of bacterial pneumonia, 17 images of *Pneumocystis* pneumonia, 23 images of fungal pneumonia, 2 images of *Chlamydophila* pneumonia, and 11 images of unidentified pneumonia. Additional 517 cases of COVID-19 pneumonia are sourced from a collaborative effort [24] (https://github.com/ieee8023/covid-chestxray-dataset). Lung segmentations in this dataset were performed by human annotators and include most of the heart, revealing lung opacities behind the heart that may be relevant for assessing the severity of viral pneumonia. The lower-most part of the lungs is defined by the extent of the diaphragm, where visible. If the back of the lungs is clearly visible through the diaphragm, it is also included up until the lower-most visible part of the lungs. Uniformly shaped lungs also decouple the shape and content within the left lung from the size of the heart. Image resolutions, sources, and orientations vary across the dataset, with the largest image being 5600x4700 and the smallest being 156x156.

### 1.2 Image pre-processing

*1.2.1 Image resizing and ground truth labels.* In order to be compatible with the image format used locally for archiving CXRs, and to facilitate further processing by our NN, all images from the listed databases and the corresponding lung masks were resized to 300 x 340 pixels. All CXRs were histogram equalized to minimize differences in contrast/brightness among the datasets and within datasets. All ground truth masks of the lung regions in these CXRs were obtained from the sources listed above. When only a semantic mask of the lung region was available, a complementary mask of the non-lung regions was generated with MATLAB. At the end of this pre-processing step, all images and ground truth binary masks were of dimensions (300 x 340 x 1), with 2 masks (one for each class, lung and non-lung) per image.

*1.2.2 Training and validation datasets.* CXRs and corresponding lung segmentations sourced from the JSRT, MC, SH, and V7-Darwin datasets were further combined into two distinct training and validation sets:

1. JMS dataset. This dataset consists of 952 CXRs derived by combining the JSTR, MC, and SH datasets. Lung masks in this dataset exclude the heart and large vasa contours. The dataset was split into a training and a validation set, with 904 and 48 image/mask pairs, respectively.

2. V7 dataset. This dataset was derived from the original V7-Darwin dataset by removing all sagittal views and CT scans. It consists of 6395 CXRs, whose corresponding lung masks include the heart and large vasa contours. The dataset was split into a training and a validation set, with 6191 and 204 image/mask pairs, respectively.

In all cases in which patient identification was available, and multiple CXRs from the same patient were included in the database, the selection of samples for the training and validation sets was carried out by carefully avoiding any *leakage* (inclusion of CXRs from the same patient in both the training and validation set).

*1.2.3 Data augmentation.* To avoid overfitting, we relied on geometric data augmentation. Each pair of image and ground truth mask(s) was sheered or rotated at random angles, shifted with a random center, vertically or horizontally mirrored, and randomly scaled in/out. The parts of the image left vacant after the transformation were filled in with reflecting padding. During training, images in a batch were not shuffled, but each image underwent a different type of augmentation as determined by the consecutive calls of a random number generator starting from a fixed initial seed. The same type of augmentation was applied to an image and its segmentation masks.

*1.3 Architecture of Res-CR-Net*

A flowchart of the architecture of Res-CR-Net is shown in **Fig. 1** (reprinted without changes from Abdallah *et al.* [22], https://doi.org/10.1088/2632-2153/aba8e8, under Creative Commons Attribution 4.0 license, https://creativecommons.org/licenses/by/4.0/legalcode ). It combines two types of residual blocks (CONV RES and LSTM RES) that are repeated in a linear path along which the dimensions of the intermediate feature maps remain identical to those of the input image and of the output mask(s).

1. CONV RES BLOCK. The residual path of this block consists of three parallel branches of separable/atrous convolutions that produce feature maps with the same spatial dimensions as the original image. Parallel branches inside the residual block are concatenated before adding them to the shortcut connection. A Spatial Dropout layer follows each residual block. In this study we have used 1 STEM BLOCK (see Fig. 1 for this block definition) and 4 CONV RES BLOCKS, each with kernel sizes of [3,3], [5,5], [7,7], and dilation rates of [1,1], [3,3], [5,5], respectively, with 16 filters in each residual branch, and 48 filters in the shortcut branch.

2. LSTM RES BLOCK. This block features a residual path with two orthogonal bidirectional 2D convolutional Long Short Term Memory (LSTM) [25] layers processing, respectively, the rows and columns of the input map from previous layers (see [22] for additional details). *m* LSTM RES BLOCKS can be concatenated. In this study we have omitted the final LSTM RES BLOCK (*m* = 0).

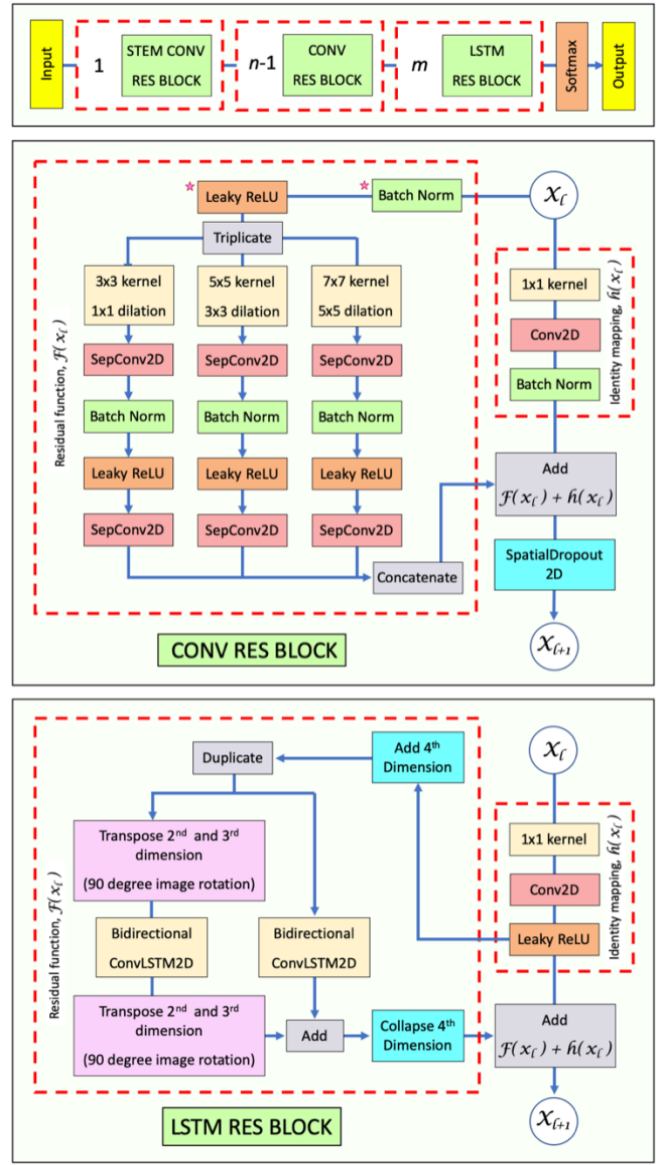

**Fig. 1.** Architecture of Res-CR-Net. In an $n+m$ levels network, the net inputs are batches of images of dimensions [batch size, rows, columns, number of channels], there are 1 STEM BLOCK, $n$-1 CONV RES BLOCKS, $m$ LSTM RES BLOCKS, and the net outputs, matching the labels, are batches of dimensions [batch size, rows, columns, number of classes] (top panel). The middle and bottom panels present a schematic view of a CONV RES BLOCK and a LSTM RES BLOCK, respectively. In this example, the parallel branches of the CONV RES BLOCK feature atrous, separable convolutions with kernels of size [3,3], [5,5], [7,7], and dilation rates [1,1], [3,3], [5,5], respectively. The STEM BLOCK differs from the CONV RES BLOCK only for lacking the operations marked with a star symbol in the CONV RES BLOCK. In the LSTM RES BLOCK, the 4D tensor emerging from the previous layer is expanded to a 5D tensor, and the following convolutional LSTM layers processes tensor slices derived either from the $2^{nd}$ and $4^{th}$ or from the $3^{rd}$ and $4^{th}$ dimension of the input feature map. This operation corresponds to scanning row-by-row both the input map and the map rotated by 90 degrees.

A *Leaky ReLU* activation is used throughout Res-CR-Net. We have not noticed any improvement in segmentation accuracy by using the *ELU* activation, as reported by Novikoff *et al.* [26]. After the last residual block a *softmax* activation layer is used to project the feature map into the desired segmentation.

The *Dice coefficient, D,* [27-31], and the *Dice loss, $L_D$*, defined as:

$$L_D(\hat{\mathbf{y}}, \mathbf{y}) = 1 - D = 1 - \frac{2\sum_i \hat{y}_i y_i + s}{\sum_i \hat{y}_i + \sum_i y_i + s}$$

with $\hat{\mathbf{y}} \equiv \{\hat{y}_i\}$, $\hat{y}_i \in [0,1]$ being the probabilities for the *i*-th pixel, $\mathbf{y} \equiv \{y_i\}$, $y_i \in \{0,1\}$ being the corresponding ground truth labels, and *s* a smoothing scalar, are often used for semantic segmentation tasks [32]. Loss functions of the *Dice* class containing squares in the denominator typically behave better in pointing to the ground truth, and even faster training convergence can be obtained by complementing the loss with a dual form that measures the overlap area of the complement of the regions of interest. These additional gains are implemented in the *Tanimoto loss*, defined as:

$$L_{\tilde{T}}(\hat{\mathbf{y}}, \mathbf{y}) = 1 - \tilde{T}(\hat{\mathbf{y}}, \mathbf{y})$$

where $\tilde{T}(\hat{\mathbf{y}}, \mathbf{y})$ is the *Tanimoto coefficient with complement*:

$$\tilde{T}(\hat{\mathbf{y}}, \mathbf{y}) = \frac{T(\hat{\mathbf{y}}, \mathbf{y}) + T(1 - \hat{\mathbf{y}}, 1 - \mathbf{y})}{2}$$

with $T(\hat{\mathbf{y}}, \mathbf{y})$ defined as:

$$T(\hat{\mathbf{y}}, \mathbf{y}) = \frac{\sum_i \hat{y}_i y_i + s}{\sum_i (\hat{y}_i^2 + y_i^2) - \sum_i \hat{y}_i y_i + s}$$

In this study we have used a *weighted Tanimoto loss* function throughout for training. Weights were derived with a *contour aware scheme*, by replacing a step-shaped cutoff at the edges of the mask foreground with a raised border that separates touching objects of the same or different classes [16]. The *unweighted Dice coefficient* defined above was used as the metric to evaluate segmentation accuracy.

### 1.4 Software

Res-CR-Net was implemented using *Keras* [33, 34] deep learning library running on top of *TensorFlow* 2.1 [35], and is publicly available at https://github.com/dgattiwsu/Res-CR-Net. Training and testing were conducted at the High Performance Computing Grid of Wayne State University.

## 2. Results

### 2.1 JMS dataset

Res-CR-Net was trained for 300 epochs with this dataset. Each epoch consisted of 113 batches of 8 images each. Thus, in every epoch the network trained on 904 different augmented images, and corresponding augmented masks as labels. Upon training Res-CR-Net achieved ~97% segmentation accuracy on the 48 images of the validation set. The training history showed no overfitting of the training set *vs.* the validation set (**Fig. 2**).

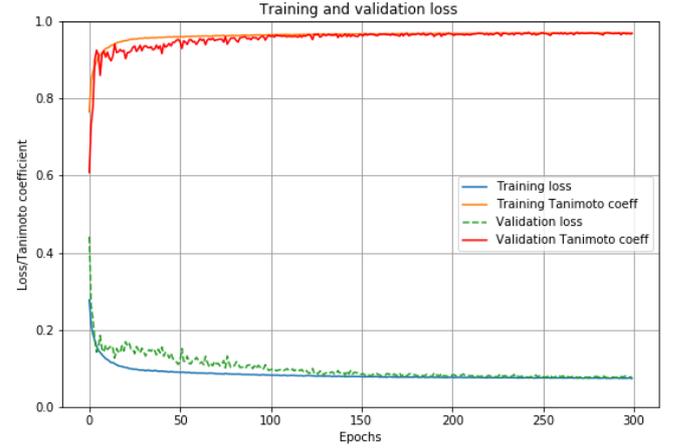

**Fig. 2.** Training and validation *weighted Tanimoto Loss* and *Tanimoto coefficient* vs. epochs for Res-CR-Net processing of JMS images.

The segmentation task in these CXR images was to identify the regions occupied by the lungs with exclusion of other thoracic structures (skeletal and cardiovascular components) but including opacities due to the underlying pathology (**Fig. 3**). The total number of parameters refined was 59,165. The execution time was 2s/batch running on 2 V100-SXM2-16GB GPUs, and the memory use was 9.8 GB/batch.

### 2.2 V7 dataset

Res-CR-Net was trained for 100 epochs with this dataset. Each epoch consisted of 516 batches of 12 images each. Thus, in every epoch, the network trained on 6192 different augmented images, and the corresponding augmented masks as labels. Upon training, Res-CR-Net achieved ~94% segmentation accuracy on the 204 images of the validation set. As observed for the JMS dataset, the training history of Res-CR-Net with the V7 dataset showed no overfitting of the training set *vs.* the validation set (**Fig. 4**).

The segmentation task in these images was to identify the regions occupied by the lungs with exclusion of the skeletal structures visible in the CXR, but including cardiovascular components and opacities due to the underlying pathology

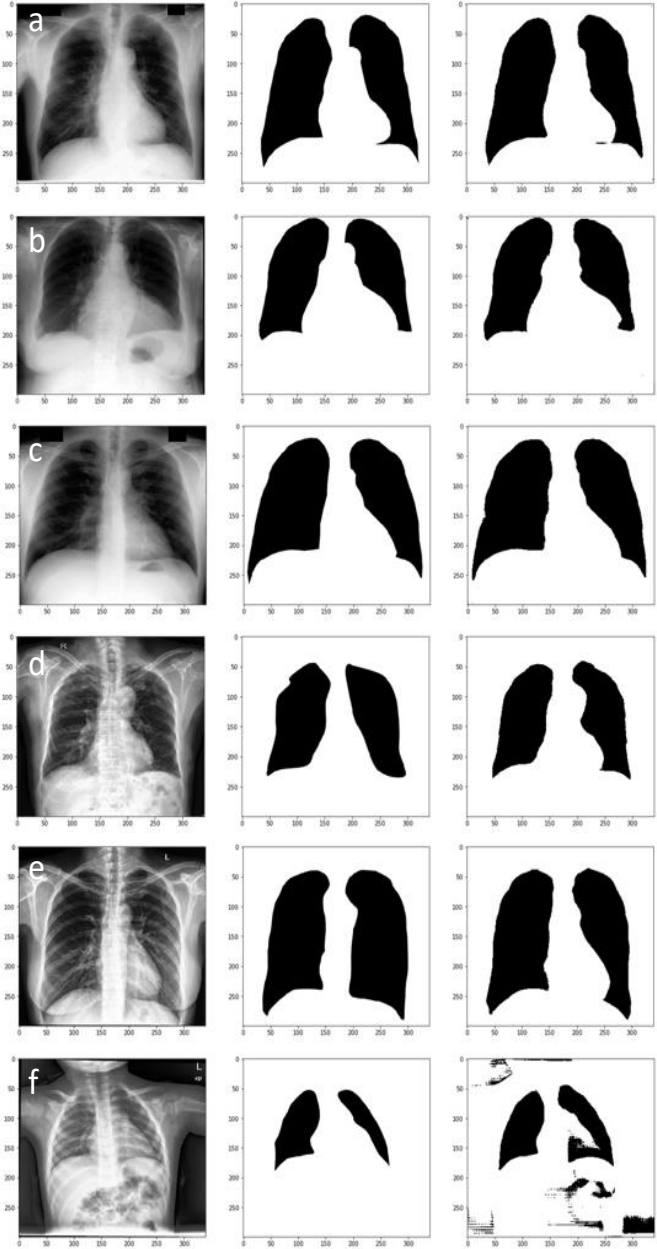

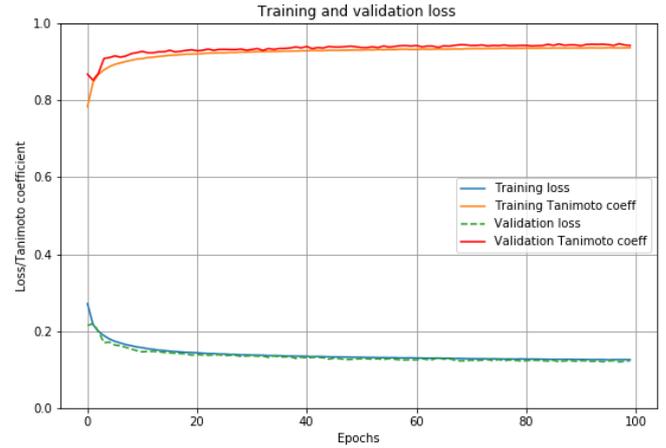

**Fig. 3.** Examples of lung segmentation in CXR images from the JMS validation subset. In each row of images: *Left panel*, CXR, *Center panel*, ground truth mask, *Right panel*, Res-CR-Net predicted mask. *Rows a-c* show three cases in which the mask is very similar to the ground truth mask. *Row d* shows an example in which the predicted mask is arguably more accurate than the ground truth mask. *Row e* shows an example in which the predicted mask excluded the heart from the lung segmentation, despite its incorrect inclusion by the annotator in the ground truth mask. *Row f* shows an example in which the predicted mask erroneously includes areas of the image that are either in the background or in the abdomen.

(**Fig. 5**). The total number of parameters refined was 59,165. The execution time was 3s/batch running on 2 V100-SXM2-16GB GPUs, and the memory use was 14.7 GB/batch.

**Fig. 4.** Training and validation *weighted Tanimoto Loss* and *Tanimoto coefficient* vs. epochs for Res-CR-Net processing of V7 images.

## 3. Conclusions

In this report, we show that Res-CR-Net, a neural network featuring a novel FCN architecture that departs significantly from the encoder-decoder paradigm, and which was originally designed for the semantic segmentation of microscopy images, also exhibits very good performance in the task of lung segmentation in CXRs from four different databases. In many cases, Res-CR-Net was effective in achieving a semantic segmentation of the lungs that was almost indistinguishable from the ground truth masks produced by human annotation (**Figs. 3,5**).

The Res-CR-Net architecture offers some advantages with respect to an encoder-decoder architecture, as its layers contain no pooling or up-sampling operations, and therefore the spatial dimensions of the feature maps at each layer remain unchanged with respect to those of the input images and of the segmentation masks used as labels or predicted by the network. For this reason, Res-CR-Net is completely modular, with residual blocks that can be proliferated in a straight down linear fashion as needed (**Fig. 1**), and it can process images of any size and shape without changing layers size and operations. In its original formulation for microscopy, Res-CR-Net also featured a novel type of residual LSTM block, in which two orthogonal convolutional LSTM layers process independently the rows and columns of the feature map emerging from the previous layers. Addition of this block to the network typically produces an additional improvement of the final segmentation masks, an effect comparable to that achieved by the conditional random field (CRF) post-refinement in other types of networks [36, 37]. In the case of lung segmentation, we found that addition of an LSTM block slowed down training with only marginal improvements in segmentation accuracy, and thus we have not included this block in the final configuration.

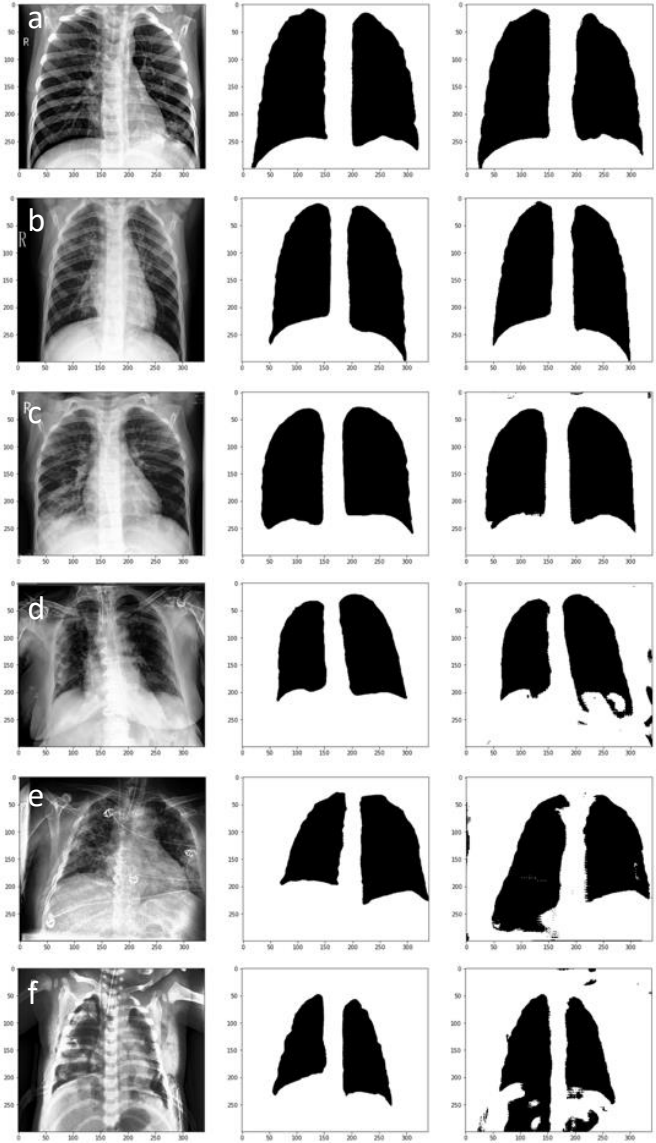

**Fig. 5.** Examples of lung segmentation in CXR images from the V7 validation subset. In each row of images: *Left panel*, CXR, *Center panel*, ground truth mask, *Right panel*, Res-CR-Net predicted mask. *Rows a-c* show three cases in which the mask is very similar to the ground truth mask. *Row d* shows an example in which the predicted mask is arguably more accurate than the ground truth mask. *Rows e-f* show two examples in which the predicted masks include areas of the images that belong to the abdomen.

The performance of Res-CR-Net with the JMS and V7 datasets is summarized in **Table 1** with respect to the metrics (*Dice coefficient, Precision, Recall, F1 score*) most often used to evaluate neural networks performance:

**Table 1.** Performance metrics for Res-CR-Net with the JMS and V7 validation datasets.

|  | [a]Dice | [b]Precision | [c]Recall | [d]F1 |
|---|---|---|---|---|
| *JMS* | 0.98 | 0.98 | 0.98 | 0.98 |
| *V7* | 0.96 | 0.96 | 0.96 | 0.96 |

[a]$Dice = \frac{2*TP}{(2*TP+FP+FN)}$,
[b]$Precision = \frac{TP}{(TP+FP)}$,
[c]$Recall = \frac{TP}{(TP+FN)}$,
[d]$F1\ score = \frac{2*precision*recall}{precision+recall}$

with TP=*true positive*, FP=*false positive*, FN=*false negative*.

Automated segmentation of the lungs is usually considered a particularly difficult task when patients have pathologies that produce lung density abnormalities that decrease the contrast between lungs and non-lungs, or when such abnormalities are superimposed to the cardiovascular structures [4]. This becomes a particularly important issue when the final goal is the unsupervised recognition of the underlying pathology based on the density features of the lung tissue, as in the end these features tend to be excluded from the predicted masks leaving only the healthy regions of the lungs. Several different strategies have been reported to achieve this goal of correctly retaining abnormal tissue in the predicted lung masks. For example, Carvahlo Souza *et al.* [10] have used a separate NN to reconstruct the dense abnormalities that had been originally ignored, and had resulted in a considerable loss of segmented lung regions. Tang *et al.* [38] have used synthetic abnormal CXRs derived from adversarial training to train their segmentation model. More recently, Selvan *et al.* [11] have treated opacifications in CXRs as missing data to be inferred, and used a variational encoder to concatenate samples from the latent space to a standard CNN for segmentation.

Here, we have shown that Res-CR-Net can be trained to generate accurate masks of the lung parenchyma that either overlap (V7 dataset, **Fig. 5**) or do not overlap the heart and large vasa (JMS dataset, **Fig. 3**), without additional modification of architecture or parameters from the configuration that had been originally optimized for the semantic segmentation of microscopy images. In that setting, Res-CR-Net was very effective at multi-class segmentation. Thus, while in this study we were only interested in lung segmentation, we anticipate that a comparable performance will be achieved also in multi-class segmentations (i.e., lungs, heart, vasa, claviculae). Furthermore, since Res-CR-Net can process images of any size, without constrains imposed by the down-pooling and up-sampling operations of an encoder-decoder architecture, it is ideally suited to be included as a pre-trained module for *on the fly* segmentation of input CXRs in a classification network that seeks to identify lung pathologies.

**Software**

Source code for Res-CR-Net is deposited at: https://github.com/dgattiwsu/Res-CR-Net

**ORCID**

Domenico Gatti: 0000-0002-6357-3530
Hassan Abdallah: 0000-0003-1256-0446

**Acknowledgements**

This study was supported by the WSU President's Research Enhancement Program in Computational Biology (DLG).